\documentclass[a4paper,fleqn,usenatbib]{mnras}
\usepackage[T1]{fontenc}
\usepackage{ae,aecompl}
\usepackage{graphicx} 
\usepackage{amsmath}  
\usepackage{amssymb}    
\usepackage{color,soul}

\title[Gas belt in radio galaxy 3C\,386]{Buoyancy-driven inflow to a relic cold core: the
  gas belt in radio galaxy 3C\,386}

\author[R. T. Duffy et al.]
{R. T. Duffy$^{1}$, 
D. M. Worrall$^{1}$, 
M. Birkinshaw$^{1}$
and R.P.~Kraft$^{2}$\\
$^1$HH Wills Physics Laboratory, University of Bristol, Tyndall Avenue,
Bristol, BS8 1TL \\
$^2$Harvard-Smithsonian Center for Astrophysics, 
60 Garden Street, 
Cambridge, MA\ \ 02138, USA \\
}

\begin{document}

\label{firstpage}

\maketitle

\begin{abstract}

We report measurements from an \emph{XMM-Newton} observation of the
low-excitation radio galaxy 3C\,386. The study focusses on an
X-ray-emitting gas belt, which lies between and orthogonal to the
radio lobes of 3C\,386 and has a mean temperature of $0.94\pm0.05$
keV, cooler than the extended group atmosphere. The gas in the belt
shows 
temperature structure with material closer to the surrounding medium being hotter than gas closer to the host galaxy. We suggest that this gas belt involves a `buoyancy-driven inflow' of part of the group-gas atmosphere where the buoyant rise of the radio lobes through the ambient medium has directed an inflow towards the relic cold core of the group. Inverse-Compton emission from the radio lobes is detected at a level consistent with a slight suppression of the magnetic field
below the equipartition value.

\end{abstract}


\section{Introduction}
Radio galaxies fuelled by active galactic nuclei (AGN) inject energy
into the surrounding medium and are important sources of heating in
galaxy groups and clusters \citep{1974Scheuer}. Energy transfer via an
AGN feedback process is inferred most directly through gas cavities
created by jet structures, which are seen frequently at the centres of
clusters in the local Universe \citep{2007McNamara}. Cavities are
formed when a radio jet does work on gas, pushing it away from the
densest regions of the group or cluster environment. Along with
cavities, many galaxies in group environments display prominent
belt-like gas structures, which are associated with the central
narrowing of radio sources between their radio lobes
\citep{2013Mannering,2013MannT}.

Belted sources such as 3C\,35, 3C\,285 and 3C\,442A have previously
been studied, with different conclusions as to their origins. For 3C\,35, \citet{2013Mannering} interpret the gas belt
as fossil-group gas driven outwards by the expanding radio lobes.
\citet{2007Hardcastle} refer to the thermal gas which is aligned
orthogonal to the radio lobes in 3C\,285 as a ridge.  They suggest the
ridge was present long before radio activity commenced, and only a
small fraction of the gas was contributed by merger activity involving
3C\,285's host galaxy.  In the case of 3C\,442A, \citet{2007Worrall}
conclude that an active merger is causing the gas of merging galaxies
to align orthogonal to, and do work on, the radio lobes.

For the three belted sources studied previously in detail, the gas appears to be either relatively stagnant or to be flowing away from the host galaxy. This could cause a barrier to positive feedback which requires AGN activity and black hole growth to be linked to gas-cooling and star formation, although the mechanisms may assist a negative feedback cycle. Alternatively, it is possible that some belts could result from gas driven inwards towards the AGN. This could potentially cause runaway positive feedback, if gas cooling from the outer atmosphere can reach the pc-scale regions around the central black hole, and provide extra fuel to renew or prolong AGN activity. In this paper, we add a fourth source, 3C\,386, to the detailed study of belted sources, using new data from \emph{XMM-Newton}. We present evidence that 3C\,386 does indeed exhibit behaviour that is opposite to that seen in the previously studied cases.

\subsection{3C\,386}

3C\,386 is a low-excitation radio galaxy with an elliptical host which
lies around 75 Mpc ($z$=0.0177) away. It is a `fat' or `relaxed double'
radio galaxy, and provides an example where the absence of jets and
hot spots \citep{1978Strom} probably indicates little current
relativistic particle acceleration \citep{2005Young}. It is a fairly
typical relaxed double source, with its lobes containing fine
structure including filaments near the edges which may form shells
\citep{1991Leahy}. It is roughly 210\arcsec\,$\times$ 290\arcsec \,in
angular size, which corresponds to projected dimensions of
76 kpc $\times$ 105 kpc in the source rest
frame. The central region of the host galaxy of 3C\,386 appears bright
in the optical due to the chance superposition of an F7 type Galactic
star \citep{1971Lynds,2009Buttiglione}. There are several other
optical point sources discernible within a radius of about 30 kpc;
some of which may be globular clusters \citep{2006Madrid}.

3C\,386's environment is classified as isolated: only two
L$^{*}$ galaxies are seen within an 800$h_{75}^{-1}$ kpc radius compared to an expected 0.9 field
galaxies \citep{1999Miller}. \citet{2013MannT} examined 3C\,386's environment using \emph{Spitzer} data taken with the Infrared Array Camera (IRAC) and found 13 faint, but plausibly associated galaxies within a 0.5 Mpc radius. Additionally, there are two galaxies with velocities within 500 km s$^{-1}$ of 3C\,386's 
velocity \mbox{\citep{2002Miller}}.  A galaxy 
close in projected distance is two magnitudes fainter than 3C\,386's host at 3.6 $\mu$m.
\citet{2013MannT} confirms the isolated environmental classification,
as potential companions are too small to contribute significant gas and/or too distant 
to cause major disturbances in 3C\,386's gas belt.

Non-detections of 3C\,386 in the X-ray were reported for both Einstein
and ROSAT \citep{1983Feigelson,1999Miller}, but it was subsequently
detected with \emph{Chandra} \citep{2010Ogle,2013MannT}. While
\citet{2010Ogle} studied the nucleus, \citet{2013MannT} investigated
several X-ray components associated with 3C\,386. Using a 30 ks
\emph{Chandra} observation (Obs ID 10232), a rectangular region
defining the gas belt indicated a temperature of $kT \approx 1.1$
keV. \citet{2015Ineson} fitted a single component radial profile through the belt and beyond and spectrally fitted all emission in an annulus of radii 2.46\arcsec to 270\arcsec to $kT = 1.05^{+0.18}_{-0.12}$. \citet{2013MannT} suggests that the gas belt is consistent with a hot
gas halo surrounding an isolated field elliptical. She finds the belt
to be overpressured relative to the surrounding environment by up to
an order of magnitude, and notes that its asymmetric morphology
suggests an interaction with the radio source in the past. She,
however, comes to no conclusion regarding the origins of the gas in
the belt.


We use a new 92 ks \emph{XMM-Newton} observation to investigate the
gas belt in 3C\,386. In Section 2, the observation and the process of
data reduction are described. In Sections 3 and 4 the X-ray morphology
of 3C\,386, and the results of the spectral fitting are
discussed. Throughout this paper a flat $\Lambda$CDM cosmology with
$\Omega_{m0} = 0.3$ and $\Omega_{\Lambda0} = 0.7$ is adopted, with
$H_{0} = 70$ km s$^{-1}$ Mpc$^{-1}$.

\section{XMM-Newton Observation and Data Reduction}

3C\,386 was observed with \emph{XMM-Newton} for 92 ks on 11-12 September
2013. Data presented here are from the EPIC MOS1, MOS2 and pn cameras in
full-frame mode. The Observation Data Format (ODF) files were
reprocessed using the tasks {\sc emchain} for the MOS detectors and
{\sc epchain} for the pn detector. We make use of the \emph{XMM-Newton} Extended Source Analysis Software procedure ({\sc XMM-ESAS}) along with the
relevant current calibration files (CCF) which contain filter wheel
closed (FWC), quiescent particle background (QPB) and soft proton (SP)
calibration data.

Due to the high variability of solar flares, the SP component of the
observation cannot be well removed using background
subtraction. Instead, frames which are dominated by high flare count
rates must be excluded, reducing the net exposure time. We created a
Good Time Intervals (GTI) file for using the {\sc XMM-ESAS} tasks {\sc
  mos-filter} and {\sc pn-filter} to generate SP
contamination-filtered products for the field of view data. The energy range used in determining the GTI for both the MOS and PN detectors was 2.5-12 keV. From the
generated light curves, it was clear that only a small part of the
observation was dominated by high flaring. Net exposure times for each
detector are listed in Table \ref{table:exposures}.

\begin{table}
	\centering
	\begin{tabular}{ccc}
	\hline
	Detector & Duration (ks) & Net exposure (ks) \\
	\hline
	MOS1 & 90.7 & 73.3 \\
	MOS2 & 90.6 & 73.8 \\
	pn & 88.1 & 60.7 \\
	\hline
	\end{tabular}
	\caption{\emph{XMM-Newton} exposure times. The duration refers to the total
          observation time, whereas the net exposure is the time
          remaining after the high flare-count-rate frames have
          been excluded.}
	\label{table:exposures}
\end{table}


For the MOS1 detector, CCD 3 and CCD 6 were excluded from subsequent
processing, as both had previously suffered from micrometeorite
damage. In the same strike which damaged CCD 3, CCD 4 sustained
collateral damage causing low energy events to dominate towards the
high $x$ values in detector coordinates (DETX). To avoid entirely
excluding CCD 4 from subsequent analysis, we created a histogram of
the number of events as a function of the $x$ position in DETX, to
determine a cutoff for good data (in this instance DETX $\approx$
9750), and included this in later spectral analysis \citep{2011Snowden}.

We used the {\sc XMM-ESAS} task {\sc cheese}, which runs source
detection on full-field images and then outputs a mask image of these
sources for use in creating source-excluded spectra. It was noted that
a few compact sources apparent by eye were not detected by the {\sc
  cheese} task, so we modified the mask to exclude them too. Following
this, the tasks {\sc mos-spectra} and {\sc pn-spectra} were run. These
extract spectra from the cleaned event files for a selected
region. Finally, the {\sc XMM-ESAS} processes {\sc mos$\_$back} and
{\sc pn$\_$back} were utilised to create particle background spectra
for the relevant regions. The \emph{XMM-Newton} background was sampled
away from the source and fitted to models using methods described in
Section \ref{sec:bkgfit}.  Spectra were binned using a minimum of 25
counts per bin, over a range of 0.4-8.5 keV for the MOS detectors and
0.4-7.2 keV for the pn detector.  A Galactic absorption of $N_{\rm H}
= 1.81 \times 10^{21}$ cm$^{-2}$ was applied to all spectral
components described as `absorbed'.

\section{X-ray Morphology}

\begin{figure}
	\centering
	\includegraphics[width=0.99\columnwidth]{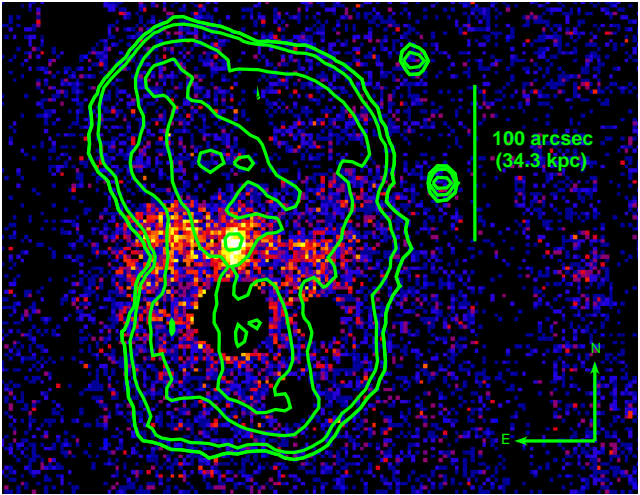}
	\caption{Particle background subtracted and exposure corrected
          MOS and pn combined \emph{XMM-Newton} image between 0.4 and 8.5 keV
          with 2.5\arcsec~pixels. Point sources excluding the AGN
          nucleus have been excised from the image. Contours are of a
          1.4 GHz VLA map 0.2, 1, 5, 10, 15 mJy/beam (5.8\arcsec\,
          beam). The radio map is taken from \citet{1991Leahy}.}
	\label{fig:combwcore}
\end{figure}

Figure \ref{fig:combwcore} shows an image of the MOS1, MOS2 and pn
data combined using the task {\sc comb}, which adds the counts from
the three instruments, accounting for differing exposures. The image
in Figure \ref{fig:combwcore} was background subtracted, to remove the
SP and QPB components. Point sources unrelated to the source are
excised.  Across the centre of the radio source, between the north and
south lobes, there is a clear excess of X-ray emission which we
describe as the gas belt. The brightening in the centre of this belt
corresponds to a superposition of the AGN core and the Galactic F7
star. The slight misalignment between these produces the elliptical
shape of the core seen in Figure \ref{fig:combwcore}. The gas belt
extends about 145\arcsec, which corresponds to a projected length of
53 kpc in the source rest frame.

\begin{figure}
	\centering
	\includegraphics[width=0.99\columnwidth]{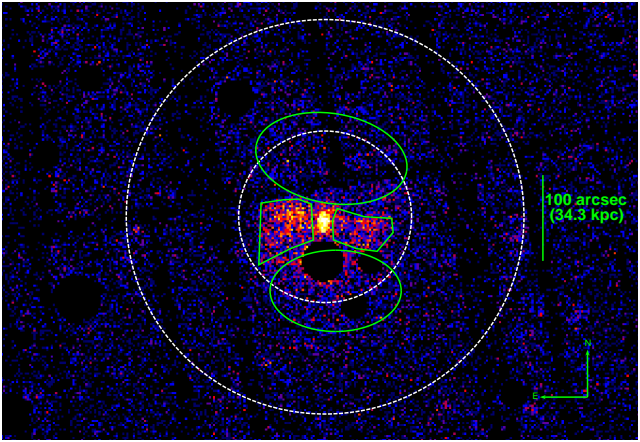}
	\caption{Image of 3C\,386 with native 2.5\arcsec~pixels, with green  regions highlighting the extraction region for the gas belt and the north and south lobes and the dashed white region highlighting the extraction region for the extended group emission.}
	\label{fig:mos2beltreg}
\end{figure}

A polygonal region for the spectral extraction across the belt was
defined by eye (see Figure \ref{fig:mos2beltreg}). The core was not
included in the spectral extraction for the belt.

\section{Spectroscopy}
\subsection{Background Estimation}
\label{sec:bkgfit}

We reduced the flare-free \emph{XMM-Newton} data using the {\sc
  XMM-ESAS} software package. For our primary analysis we chose to
model the background, using local background subtraction to check for
consistency. The model background used spectra extracted from a
circular off-source region, away from the source and its associated
gas belt and group gas. The location of the extraction region differs
between detectors in order to avoid chip gaps, but for each it is
extracted from an area of blank sky. The local background region was
extracted from an annulus surrounding 3C\,386.

The modelled background is preferred to subtracting a local background
as it can accommodate variation in background counts across the field
of view. Due to this, unless explicitly stated, values quoted in this
paper will be those measured from fits with modelled background. The
local background is simply used as a method of cross checking.

We follow the prescription of \citet{2013Mannering}, who in turn
followed \citet{2011Snowden}, to model the \emph{XMM-Newton}
background component. The \emph{XMM-Newton} tasks {\sc mos$\_$back} and {\sc pn$\_$back} generate model QPB spectra from the FWC observations and
unexposed corners of the MOS and pn CCDs, respectively. This quiescent
instrumental background is subtracted from data, and then additional
components in the \emph{XMM-Newton} background need to be modelled
explicitly. Instrumental Al K$\alpha$ and Si K$\alpha$ lines were
fitted with unabsorbed Gaussians of zero intrinsic widths at line
energies of 1.49 keV and 1.75 keV, respectively, for the MOS
detectors, whilst only the Al K$\alpha$ line at 1.49 keV was required
for the pn data. The energy range for the pn detector was restricted
between 0.4 and 7.2 keV, due to the difficulty in modelling the QPB at
high energies for this detector and the large number of instrumental
Cu K$\alpha$ lines present at 7.49 keV and above.

The cosmic X-ray background was modelled by an unabsorbed thermal
component representing emission from the heliosphere at $kT \approx
0.1$ keV, a higher temperature ($kT \approx 0.25 - 0.7$ keV) absorbed
thermal component representing emission from the hotter halo and
intergalactic medium, and an absorbed power law with $\Gamma \approx
1.46$ representing the unresolved background of cosmological
sources. We also included a cool, ($kT \approx 0.1$ keV) absorbed
thermal component representing emission from the cooler halo, but this
was fit with a very small normalisation and was unnecessary. In order
to further constrain the cosmic background contribution, a ROSAT All
Sky Survey (RASS) spectrum from the HEASARC X-ray Background Tool was
also included.

Solar Wind Charge Exchange (SWCX) emission, which originates in the
heliosphere, contributes a significant fraction of the X-ray
background at energies of less than 1 keV. The ions of high charge
state in the solar wind interact with neutral atoms, thereby gaining
an electron in a highly excited state. This electron decays by
emitting X-rays, which produces a background strongly dependent on the
solar wind proton flux and heavy ion abundances. We included several
SWCX lines in our \emph{XMM-Newton} background model, represented by additional Gaussians at several line energies with zero intrinsic width. These were identified by fitting a model containing the instrumental Gaussian and cosmic X-ray background components, and adding the SWCX at relevant energies to improve the fit.

\begin{figure}
	\centering
	\includegraphics[height=3.40in,angle=270]{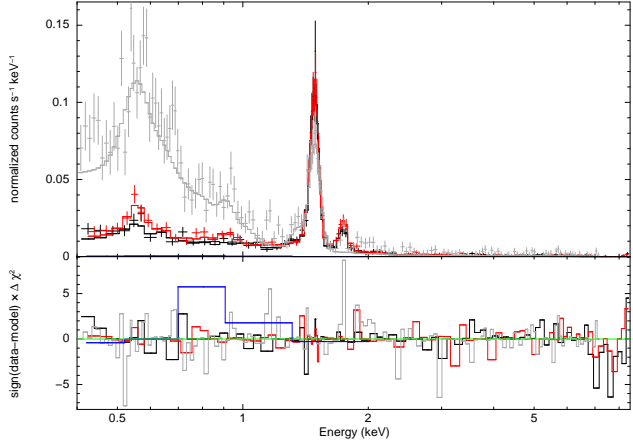}
	\caption{\emph{XMM-Newton} spectra extracted from the off-source
          region, and fit to the background model described in Section
          \ref{sec:bkgfit}. Black, red, and grey
          correspond to MOS1, MOS2, and pn data, respectively.}
	\label{fig:bkgfit}
\end{figure}

Data and responses for the three XMM detectors were not combined before spectral fitting. Rather, their data were fitted concurrently to the same models. Between each data set several parameters are held common; these include the energies and widths of each Gaussian used to describe a line, and parameters such as the redshift of the source and the abundances and temperatures of each part of the cosmic X-ray background model.
The best-fit model for the off-source data (Fig.~\ref{fig:bkgfit})
has $\chi^{2} =$ 600.4 for 621 degrees of freedom.

\begin{table*}
	\begin{tabular}{c c c c c c c}
		\hline
		& \multicolumn{3}{ c }{Spectral data counts} & \multicolumn{3}{ c }{Area (arcmin$^{2}$)} \\ 
		& MOS1 & MOS2 & pn & MOS1 & MOS2 & pn \\
		\hline
		Whole gas belt & 735 (88.6\%) & 774 (89.7\%) & 1926 (87.7\%) & 1.452 & 1.443 & 1.536 \\
		Inner belt & 373 (91.7\%) & 383 (92.4\%) & 1035 (89.9\%) & 0.663 & 0.619 & 0.716 \\
		Outer belt & 369 (87.2\%) & 413 (88.5\%) & 930 (85.0\%) & 0.809 & 0.838 & 0.839 \\
		North lobe & 848 (73.5\%) & 806 (70.8\%) & 1610 (67.0\%) & 3.870 & 3.781 & 3.273 \\
		South lobe & 740 (79.9\%) & 642 (78.5\%) & 1571 (74.0\%) & 2.558 & 2.294 & 2.542 \\
		Core & 171 (97.2\%) & 113 (95.7\%) & 525 (96.0\%) & 0.113 & 0.132 & 0.134 \\
		Group gas & 4761 (59.3\%) & 4907 (58.7\%) & 8820 (51.8\%) & 32.24 & 32.92 & 27.39 \\
		Background & 4834 (41.1\%) & 4842 (48.7\%) & 8910 (42.4\%) & 39.40 & 39.43 & 38.20 \\
		\hline
		\end{tabular}
	\caption{Properties of the regions used in spectral
          extraction. Spectral data counts are measured between 0.4-8.5 keV
          for the MOS1 and MOS2 detectors, and between 0.4-7.2 keV for
          the pn detector. The counts refer to the source,
          instrumental lines, SWCX, SP and CXB emission, minus the QPB
          background which is removed from the data prior to
          fitting. The percentages indicate the source counts
          over total counts in the region. Area is the net solid angle
          of the selected region, excluding chip gaps, damaged CCDs
          and background sources.}
	\label{table:countrates}
	\end{table*}

\subsection{Gas Belt}

The \emph{XMM-Newton} spectra were extracted from a polygonal region of the gas belt. This region lies orthogonal to the lobes of 3C\,386 and is
shown in Figure \ref{fig:mos2beltreg} superimposed on a combined image
from all detectors. Counts from the east and west regions shown in Figure \ref{fig:mos2beltreg} were extracted as a single spectrum, in order to model the gas belt as a whole. The spectral data counts for each detector are given in Table \ref{table:countrates}. The \emph{XMM-Newton} background is modelled as described in Section \ref{sec:bkgfit}. The line energies and widths of the instrumental and SWCX lines determined as in Section \ref{sec:bkgfit} were frozen for the on-source fits, with only the normalisations of the lines allowed to vary.

Details from all fits related to the gas belt are given in Table
\ref{table:beltfit}. The gas belt \emph{XMM-Newton} data fit a single
absorbed APEC plasma model \citep{2001Smith}, with a fixed metal abundance of
0.3$Z_{\odot}$, of temperature $0.94^{+0.04}_{-0.05}$ keV. The data
were also fitted with an absorbed power law, giving a reasonable but
statistically poorer fit and a photon index of $\Gamma =
2.33\pm0.11$. A combination of an absorbed APEC and power-law fit
finds a consistent gas temperature of $0.90^{+0.06}_{-0.09}$ keV. In
each case, across the entire gas belt, there is good agreement between
the parameters obtained from the model background fits and those
obtained with a subtracted local background.


\subsubsection{Gas Belt Temperature Structure}

\begin{figure}
	\centering
	\includegraphics[width=0.99\columnwidth]{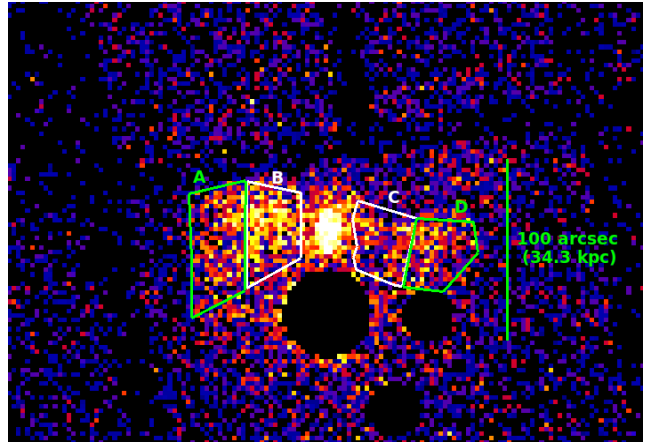}
	\caption{Image of 3C\,386 with native 2.5\arcsec~pixels,
          highlighting the extraction regions for the inner (B \& C)
          and outer (A \& D) regions of the gas belt.}
	\label{fig:abcdsplit}
\end{figure}


In an effort to establish the mechanism by which the belt was shaped,
we further divide the belt region shown in Figure
\ref{fig:mos2beltreg}, into two smaller sub-regions. The sub-regions
are shown in Figure \ref{fig:abcdsplit}, and are labelled as `inner'
and `outer' regions corresponding to their position within the
belt. Both inner and outer regions were fitted with the same models as
the whole gas belt. When fitted to an APEC model, the temperature of
the inner gas belt was found to be $0.73^{+0.08}_{-0.11}$ keV whilst
the outer belt was found to have a temperature of
$1.72^{+0.56}_{-0.36}$ keV.

If we assume the gas belt is a disk we are viewing side on, there
would be some outer belt material in our line of sight in front and
behind the inner region of the belt. To compensate for this and gain a
more realistic measurement of the temperature of the inner belt, we
fitted an APEC + APEC model to the inner belt spectra. The first APEC
has a temperature fixed to the best-fit value measured from the outer
belt fit, and its normalisation is set to accommodate the outer gas
assuming a disk geometry. The second APEC, corresponding to only the
inner material is then fitted to a temperature of
$0.62^{+0.12}_{-0.10}$ keV, somewhat lower than before. Once we 
established the temperatures of the two gas components, we
performed a second APEC + APEC fit, this time to the entire gas belt.
In this fit the temperatures were fixed at the inner and outer component values. This enabled us
better to gauge the spatial extent of each temperature component, as
discussed further in Section \ref{section:gasbeltdisc}.

Each subdivision of the belt can alternatively be fitted to an absorbed power law,
with $\Gamma = 2.70^{+0.37}_{-0.28}$ for the inner gas belt, and
$\Gamma = 2.10^{+0.17}_{-0.16}$ for the outer gas belt. A combination
of an absorbed APEC and a power law finds the temperature of the inner
gas belt $0.67^{+0.10}_{-0.09}$ keV and the outer gas belt
$1.56^{+0.41}_{-0.18}$ keV, and the power law 1 keV normalisations are
an order of magnitude smaller than those of the thermal
components. Each fit described above has a fixed metal abundance with
respect to solar of $Z_{\odot} = 0.3$. Further details can be found in
Table \ref{table:beltfit}.


\begin{figure}
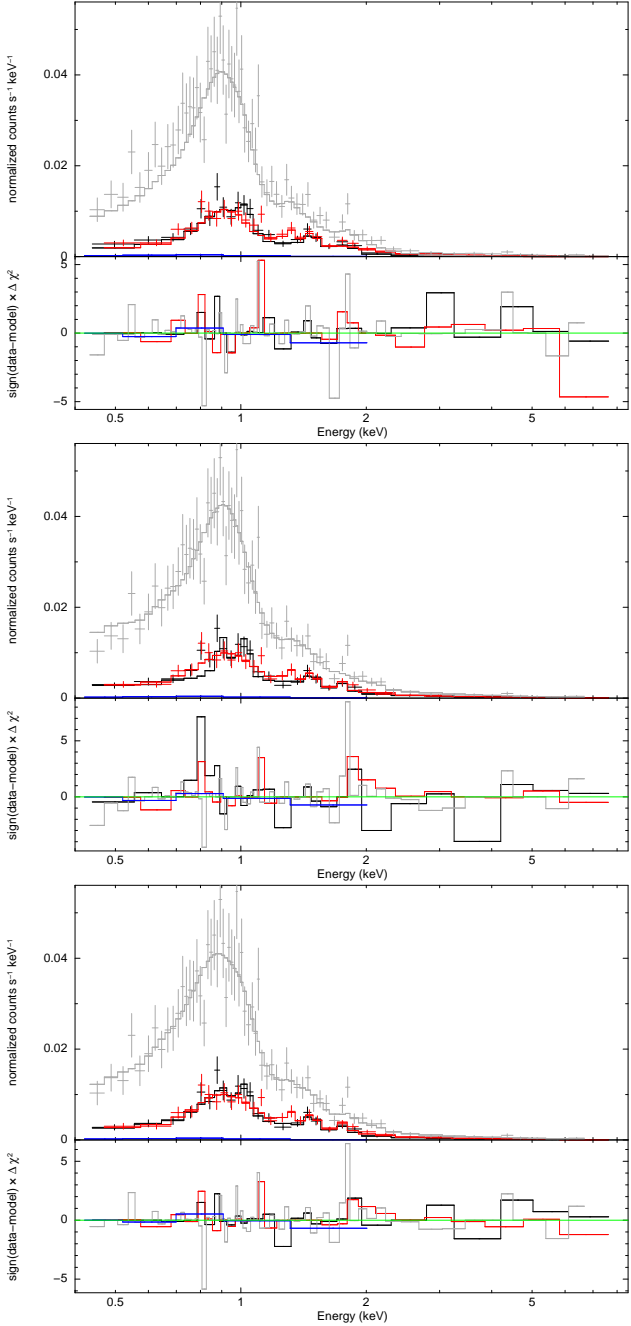

	\centering
	\includegraphics[height=3.40in,angle=270]{beltapecfit3.ps}
	\includegraphics[height=3.40in,angle=270]{beltabspofit3.ps}
	\includegraphics[height=3.40in,angle=270]{beltapecpofit3.ps}
	\caption{\emph{XMM-Newton} spectra extracted from the whole gas belt,
          fitted to the background model plus an APEC component in the
          top panel, an absorbed power law in the middle panel and an
          APEC + power-law model in the bottom panel. The lower panel
          in each fit show the $\chi^{2}$ residuals. In each panel,
          black, red, and grey
          correspond to MOS1, MOS2, and pn data, respectively.}
	\label{fig:gasbeltfits}
\end{figure}
	
We also performed a second APEC fit on the inner and outer gas belts,
this time allowing the metal abundances in each region to vary along
with the other parameters. We find no significant change in
temperature, with metal abundance of $Z_{\odot} =
0.12^{+0.12}_{-0.05}$ and $Z_{\odot} = 0.68 \pm 0.49$ in the inner and
outer regions respectively.

For the discussion, we assume that the gas belt is entirely of thermal
origin, since any power law contribution is small.

The volume of the gas belt was estimated as $49\times10^{3}$
kpc$^{3}$, by assuming it is part of a disk of radius 83\arcsec
($\sim30$ kpc) and thickness 50\arcsec ($\sim18$ kpc). The radius and thickness of the disk were estimated assuming that the disk is seen edge-on.

\begin{table*}
	\begin{tabular}{l c c c}
		\hline
		Model parameter & Whole gas belt & Inner belt & Outer belt \\
		\hline
		APEC & & & \\
		$kT$ (keV) & 0.94$^{+0.04}_{-0.05}$ & $0.73^{+0.08}_{-0.11}$ & $1.72^{+0.56}_{-0.36}$ \\
		$10^{14}N$ ($10^{9}$ cm$^{-5}$) & 3.83$^{+0.63}_{-0.62}$ & $1.78^{+0.43}_{-0.42}$ & $2.22\pm0.70$ \\
		$\chi^{2}$/dof & 95.0/98 & 75.6/82 & 70.4/80 \\
		\hline
		APEC + APEC & & & \\
		$kT_{1}$ (keV) & 1.72 (fixed) & 1.72 (fixed) & - \\
		$10^{14}N_{1}$ ($10^{9}$ cm$^{-5}$) & $3.92\pm0.56$ & 1.04 (fixed) & - \\
		$kT_{2}$ (keV) & 0.62 (fixed) & $0.62^{+0.12}_{-0.10}$ & - \\
		$10^{14}N_{2}$ ($10^{9}$ cm$^{-5}$) & $1.83\pm0.32$ & $1.01^{+0.47}_{-0.46}$ & - \\
		$\chi^{2}$/dof & 95.8/101 & 75.2/78 & - \\
		\hline
		 power law & & & \\
		$\Gamma$ & 2.33$\pm$0.11 & $2.70^{+0.37}_{-0.28}$ & $2.10^{+0.17}_{-0.16}$ \\
		$S_{\rm 1~keV}$ (nJy) & 10.49$^{+1.57}_{-1.55}$ & $5.01^{+1.37}_{-1.34}$ & $4.20^{+1.20}_{-1.14}$ \\
		$\chi^{2}$/dof & 112.6/106 & 93.0/90 & 85.5/88\\
		\hline
		APEC + power law & & & \\
		$kT$ (keV) & 0.90$^{+0.06}_{-0.10}$ & $0.67^{+0.10}_{-0.09}$ & $1.56^{+0.41}_{-0.18}$ \\
		$10^{14}N_{\rm apec}$ ($10^{9}$ cm$^{-5}$) & 3.47$^{+0.64}_{-0.39}$ & $1.78^{+0.33}_{-0.34}$ & $2.40^{+0.70}_{-0.88}$ \\
		$\Gamma$ & 1.34$^{+0.32}_{-0.39}$ & $1.24^{+0.63}_{-0.20}$ & $1.07^{+1.06}_{-0.43}$ \\
		$N_{\rm power~law}$ & 0.38$^{+0.32}_{-0.30}$ & $0.22^{+0.13}_{-0.11}$ & $0.04^{+0.26}_{-0.04}$ \\
		$(10^{-5}$cm$^{-2}$s$^{-1}$keV$^{-1})$ & & & \\
		$\chi^{2}$/dof & 84.5/95 & 94.5/79 & 75.0/77 \\
		\hline
		\end{tabular}
	\caption{Best fit parameters of the emission from across the whole gas belt, the inner gas belt and the outer gas belt of 3C\,386. Models fit are an absorbed APEC,  power law and APEC plus  power law. Metal abundances in the APEC, APEC + APEC and APEC +  power law fits are fixed at $Z_{\odot}$ = 0.3. Errors are quoted for a 90\% confidence range.}
	\label{table:beltfit}
	\end{table*}
	

\subsubsection{Physical properties of the gas belt}
\label{section:gasbeltdisc}

The gas belt of 3C\,386 is well described by a thermal model. Using
the fitted parameter values along with the estimated volume we
calculate the pressure and total mass of the gas belt following the
methods described in \citet{2006Worrall}. The emission measure,
defined in terms of the normalisation factor given by the APEC model,
$N$, is given by

\begin{align}
\centering
10^{14}N = \frac{(1 + z)^{2} \int n_{\rm e}n_{\rm p} dV}{4 \pi D_{\rm L}^{2}}
\end{align}

\noindent where $D_{\rm L}$ is the luminosity distance in cm, $V$ is the
volume in cm$^{3}$ and $n_{\rm p}$ is the proton density in
cm$^{-3}$. With normal cosmic values for element abundances, the
proton number density for a region of uniform density is given by

\begin{align}
\centering
n_{\rm p} \approx \sqrt{\frac{10^{14} N 4 \pi D_{\rm L}^{2}}{(1 + z)^{2} 1.18 V}}
\end{align}

\noindent If the proton density is constant over the volume, the
pressure can then be found by

\begin{align}
P = 3.6\times10^{-10}n_{\rm p}kT \quad\text{Pa}
\end{align}

\noindent where $n_{\rm p}$ is in units of cm$^{-3}$ and $kT$ is in
units of keV \citep{2012Worrall}. 

A summary of results for the gas belt are given in Table \ref{table:xrayprop}. 
The densities and pressures in the gas belt correspond to the average density
and average pressure over the region of the spectral extraction.
The masses calculated in Table \ref{table:xrayprop} for the inner and outer
belt were found using the APEC + APEC fit for the inner region and the APEC fit for
the outer region. The distinction between the inner and outer belt was made at
approximately the mid-point in the belt on either side of the core. We tested that this
distribution of gas is roughly correct, by fitting an APEC + APEC to the entire belt
with each APEC fixed at the inner and outer temperature. When doing this, we find a
mass for the inner belt of log$(M/M_{\odot}) = 8.8\pm0.1$ and a mass for the outer 
belt of log$(M/M_{\odot}) = 9.1\pm0.1$. These masses are similar to those in Table
\ref{table:xrayprop}, supporting our choice of the relative spatial extents of 
the inner and outer belt regions.

\begin{table}
	\centering
	\begin{tabular}{l c c c}
		\hline
		Parameter & Gas belt & Inner belt & Outer belt \\
		\hline
		$kT$ (keV) & 0.94$^{+0.04}_{-0.05}$ & $0.62^{+0.12}_{-0.10}$ & 1.72$^{+0.56}_{-0.36}$ \\
		Volume (kpc$^{3}$) & $49 \times 10^{3}$ & $13 \times 10^{3}$ & $36 \times 10^{3}$ \\
		$n_{\rm p}$ (m$^{-3}$) & $1250\pm200$ & $1200\pm100$ & $1300\pm100$ \\
		$P$ (10$^{-13}$ Pa) & $4.1\pm0.2$ & $2.8^{+0.8}_{-0.9}$ & 8.2$_{-1.8}^{+2.8}$ \\
		log($M/M_{\odot}$) & $9.2\pm0.1$ & $8.6\pm0.1$ & $9.0\pm0.1$ \\
		$\bar{r}$ (kpc) & 15 & 10 & 21 \\
		\hline
		\end{tabular}
	\caption{Temperature, density, and mass of the X-ray emission for the belt estimated from the \emph{XMM-Newton} observations. $\bar{r}$ corresponds to the average distance from the core for regions of the gas belt.}
	\label{table:xrayprop}
\end{table}

\subsection{Lobes}
\label{sec:speclobes}
The \emph{XMM-Newton} spectra for the radio lobes were extracted from
two elliptical regions north and south of the gas belt. Contaminating
sources (holes seen in Figure \ref{fig:mos2beltreg}) are excluded.  Also in Figure \ref{fig:mos2beltreg}, 
there is some excess flux visible from the wings of the PSF of an excluded 
bright GALEX source. These remaining counts contribute a small fraction 
of the total counts in the southern lobe, and have no significant impact 
on the spectral fits. The net count rates for each detector for each lobe
are given in Table \ref{table:countrates}. The
lobes were fitted separately. We performed a combined 
fit of the MOS and pn spectra. The source emission in both
lobes is best fitted by a power law with fixed Galactic
absorption. These fits find a photon index of $\Gamma =
2.02^{+0.23}_{-0.21}$ ($\alpha_{X} = 1.02^{+0.23}_{-0.21}$) for the
northern lobe and $\Gamma = 1.57^{+0.12}_{-0.11}$ ($\alpha_{X} =
0.57^{+0.12}_{-0.11}$) for the southern lobe. See
Table~\ref{table:lobefit} for details. The difference in the spectral indices of these lobes is probably caused by different amounts of foreground and background gas along the line of sight. When both lobes are fit
together we find $\alpha_{X} = 0.73^{+0.11}_{-0.07}$.
Using measurements of flux density from the lobes of 3C\,386 between
178 and 5000 MHz collected from the NASA/IPAC Extragalactic Database
(NED), we find a spectral index $\alpha_{\rm r} = 0.72\pm0.16$. The
X-ray and radio indices agree within measurement errors.
\begin{table}
	\centering
	\begin{tabular}{l c c}
		\hline
		Model parameter & North lobe & South lobe \\
		\hline
		$N_{\rm H}$ (10$^{21}$ cm$^{-2}$) & 1.81 (fixed) & 1.81 (fixed) \\
		$\Gamma$ & 2.02$^{+0.23}_{-0.21}$ & 1.57$^{+0.12}_{-0.11}$ \\
		$S_{\rm 1~keV}$ (nJy) & 7.05$^{+1.40}_{-1.39}$ & 6.28$^{+0.76}_{-0.82}$ \\
		$\chi^{2}$/dof & 227.5/268 & 295.0/310 \\
		1.4 GHz Radio flux density (Jy) & $2.97\pm0.31$ & $1.39\pm0.21$ \\
		\hline
		\end{tabular}
	\caption{Summary of the power law model fitting to the north
          and south lobes of 3C\,386, along with measured radio flux
          densities. The Galactic neutral hydrogen column density is
          frozen for all fits. Errors are quoted for a 90\% confidence range.}
	\label{table:lobefit}		
\end{table}

The lobes were also fitted to APEC models. With a fixed abundance of
0.3$Z_{\odot}$, the temperatures are found to be 3.3 $\pm$ 0.8 keV for
the northern lobe and 6.2 $\pm$ 1.3 keV for the southern lobe. These
temperatures are unrealistically high for 3C\,386, as they are more
consistent with those expected in a cluster environment. No evidence
from either the X-ray surface brightness or the galaxy's environment
suggests that 3C\,386 resides in a cluster.  We therefore conclude
that the emission in the lobes is non-thermal, even though the
$\chi^{2}$ values for the thermal fits are acceptable.

Approximating oblate ellipsoids for the northern and southern lobes,
we estimate the volumes corresponding to X-ray emission to be
$81\times10^{3}$ kpc$^{3}$ and $42\times10^{3}$ kpc$^{3}$,
respectively. The volume for the southern lobe has been corrected down
to account for the excised regions. The 1.4 GHz VLA radio flux density
was measured to be $2.97\pm0.31$ Jy for the northern lobe and
$1.39\pm0.21$ Jy for the southern lobe over the same areas used for
the X-ray spectral extraction.

\subsection{The Core}
\begin{figure}
	\centering
	\includegraphics[width=0.99\columnwidth]{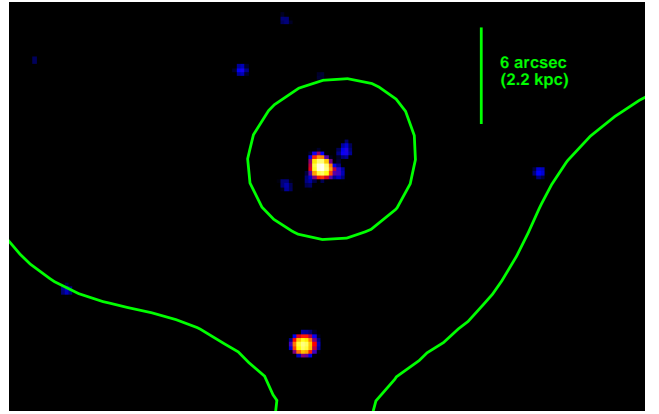}
	\caption{0.5-7.0 keV \emph{Chandra} image of the core of 3C\,386 with 0.246\arcsec pixels, smoothed with a Gaussian of standard deviation 0.37\arcsec. Contours are of a 1.4 GHz VLA map 10, 15 mJy/beam (5.8\arcsec\,beam). The northern X-ray component within the radio map's centre corresponds to the radio nucleus, while the component to the south corresponds to stellar emission. These two components are confused in the \emph{XMM-Newton} images.}
	\label{fig:corezoom}
\end{figure}
The core is a superposition of the radio nucleus and an F7-type
Galactic star. These two components can be seen in the \emph{Chandra}
observation of 3C\,386 (see Figure \ref{fig:corezoom}), with a northern component corresponding to the radio nucleus, and a separate southern component corresponding to emission from the star. Despite this, the spectrum of the core was extracted from a circular region with radius
12.5\arcsec. The spectral data from the core were fitted to an
absorbed power law with fixed Galactic absorption of $N_{\rm H} = 1.81
\times 10^{21}$ cm$^{-2}$. The best-fit photon index was found to be
$\Gamma = 2.11^{+0.26}_{-0.25}$ and the best-fit 1 keV flux density
corresponds to $S_{\rm 1~keV} = 3.01^{+0.82}_{-0.85}$ nJy, with
$\chi^{2}/\text{dof} = 126.1/133$. The spectral fit shows some
indication of structured residuals between 0.6 and 0.9 keV. There is
no evidence of excess absorption and results are consistent with the
cores of other FRI radio galaxies \citep{2006Evans}. The 2-10 keV
X-ray luminosity of the core is $6.54 \times 10^{39}$ ergs s$^{-1}$,
which is roughly consistent with the value obtained by
\citet{2010Ogle}. Our derived X-ray luminosity is an upper limit
because of contaminating emission from the star.

\subsection{Extended Group Gas}
\label{sec:gg}

\begin{figure}
	\centering
	\includegraphics[height=3.40in,angle=270]{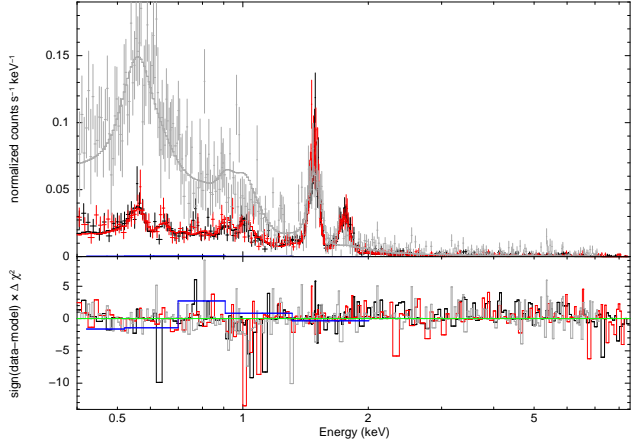}
	\caption{\emph{XMM-Newton} spectra extracted from an annulus of group
          gas lying between 100\arcsec and 230\arcsec \,from the core
          of 3C386, fitted with an APEC + background model. The lower
          panel shows the $\chi^{2}$ residuals. Black, red, and grey
          correspond to MOS1, MOS2, and pn data, respectively.}
	\label{fig:groupgasfit}
\end{figure}

Group-scale gas is evident in the \emph{XMM-Newton} data, although was
not necessarily expected given what is known of the galaxy environment
(Section 1.1).  The spectrum of this component was first extracted
from an annulus surrounding the galaxy, extending between 100\arcsec\,
to 230\arcsec \,from the core, and so beyond the belt and mostly
beyond the radio lobes. To check the consistency of our results we
also extracted a spectrum from a small rectangular region of less than
6\% of the whole area (2.1 square arcmin, 2\arcmin \,to the southwest
of the core) which had good coverage on all three EPIC detectors. The
results of the fit from the smaller rectangular region are used as a
method of testing the reliability of the fit from the larger annulus,
as there is often difficulty in consistently modelling the
\emph{XMM-Newton} background over large detector areas. The background
in both of the extracted regions was modelled as described in Section
\ref{sec:bkgfit}. The data were fitted to an APEC model. For the
annulus extraction, the temperature of the group is poorly
constrained, with the best-fit temperature too high for the X-ray
luminosity and inferred richness of the group. We have therefore fixed
the temperature at 1.4 keV, which is roughly the $1\sigma$ lower
bound. How this model fits the data for the group gas is shown in
Figure \ref{fig:groupgasfit}, with parameter values given in Table
\ref{table:gasfit}. The rectangular region produces a temperature of
1.72$^{+4.59}_{-0.94}$ keV, which is consistent with our temperature
choice for the entire annulus. Although the errors on the measured
temperatures are large, the \emph{XMM-Newton} data support the
existence of a component of extended emission of group scale and
temperature.

\begin{table}
	\centering
	\begin{tabular}{l c}
		\hline
		Model parameter & Group gas \\
		\hline
		$kT$ (keV) & 1.4 (fixed) \\
		$10^{14}N$ ($10^{9}$ cm$^{-5}$) & $5.68^{+1.68}_{-1.89}$ \\
		$\chi^{2}$/dof & 713.0/707 \\
		\hline
		\end{tabular}
	\caption{Summary of the APEC fitting to the group gas
          surrounding 3C\,386. The Galactic neutral hydrogen column
          density is frozen for all fits. Metallicity is fixed to
          0.3$Z_{\odot}$. Errors are quoted for a 90\% confidence
          range.}
	\label{table:gasfit}		
\end{table}

\begin{figure}
	\centering
	\includegraphics[width=0.99\columnwidth]{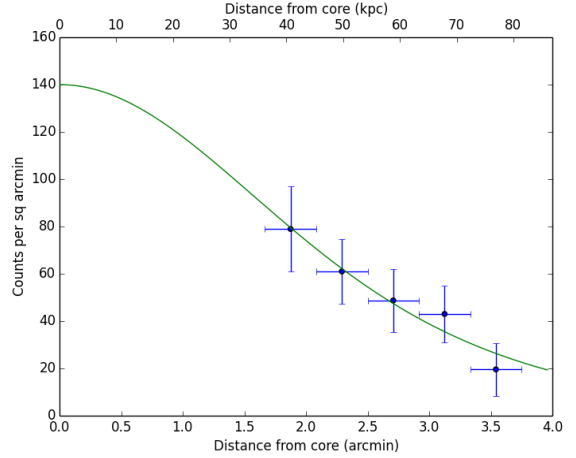}
	\caption{Vignetting-corrected, background subtracted radial
          profile of extended emission from the pn detector from ~35
          to ~80 kpc from the core of 3C\,386. Data have been filtered
          between 0.4-7.2 keV. The lobes and any additional point
          sources were masked in each annulus. The green line
          indicates a $\beta$-model where $\beta = 1.2$ and $r_{\rm cx} =
          91$ kpc.}
	\label{fig:radprof}
\end{figure}

To investigate the radial profile of the group emission, counts were
extracted from the pn detector from annuli of width 25\arcsec.  We
found a profile that decreased with increasing radius out to about 3.8
arcmin, at which point it flattened.  This was taken to indicate the
point at which the group emission falls significantly below the
background. We used this information to subtract a background from
each of the inner bins.  The lobes and any additional point sources
inside the annuli were masked so that only counts from group emission
were counted.  The resulting background-subtracted
vignetting-corrected radial profile of X-ray gas surrounding 3C\,386
is shown in Figure \ref{fig:radprof}.  

The radial profile can be a fitted with a $\beta$-model. We used $\chi^{2}$ 
fitting to establish the best fit to the model, with $\beta$ limited between 
0.4 and 1.2 and find a best fit with $\beta = 1.2^{\rm +unknown}_{-0.8}$, 
core radius, $r_{\rm cx} = 91^{+56}_{-82}$ kpc and central counts, 
$B_{o} = 140^{+460}_{-65}$. The best fit $\chi_{\rm min}^{2} = 0.8$, with 
errors on each interesting parameter calculated to 90\% confidence within 
$\chi^{2}_{\rm min} + 2.7$ allowing the other parameters to vary.
the fitted core radius is $r_{\rm cx} = 30$ kpc. This is similar to the 
length of the gas belt, and so while group gas is detected out to 80 kpc, 
the gas belt lies within its core radius.
Using equation 10 from \citet{1993Birkinshaw}, the
central proton density in cm$^{-3}$ from this model profile can be
calculated.

\begin{equation}
\eta_{po} = \frac{2.45(1 + z)^{3}}{a^{\frac{3}{2}}}
\sqrt{\frac{B_{o}f\Gamma(3\beta)}{t\theta_{\rm cx}D_{\rm L}\Gamma(3\beta-0.5)}}
\end{equation}

\noindent where $B_{o}$ is the central brightness in count
arcmin$^{-2}$, $a$ is the number of radians per arcmin, $f$ is a
calibration quantity equivalent to 1 ct/s in cm$^{-5}$ for a fitted
{\sc xspec} normalisation, $t$ is the exposure time, $\theta_{\rm cx}$ is
the core radius in arcmin and $D_{\rm L}$ is the luminosity distance in
cm.

Using this central proton density, the gas mass contained within
multiples of the core radius can be calculated using the following
equation:

\begin{equation}
M(r \leq kr_{\rm cx}) = \frac{4k^{3}\pi m_{\rm H} \eta_{\rm po} r_{\rm cx}^{3}}{X}
\int_{0}^{1} dx\, x^{2} (1 + k^{2}x^{2})^{-\frac{3\beta}{2}}
\end{equation}

\noindent where $x = r/kr_{\rm cx}$ is used as a substitution, $k$ is a
multiple of the core radius, $m_{\rm H}$ is the mass of a hydrogen atom
and $X$ is the cosmic abundance of hydrogen by mass. Using this
integral we find the gas mass contained within $r_{\rm cx}$ is $3.0^{+4.5}_{-2.9}
\times 10^{10} M_{\odot}$ and the gas mass contained within $2r_{\rm cx}$
is $8.1^{+12.0}_{-8.0} \times 10^{11} M_{\odot}$. 


\section{Discussion}


Assuming that the belt is of constant density, we estimate
the mass of gas it contains as log($M/M_{\odot}) =
9.2\pm0.1$. When calculating the belt mass, we have continued to exclude the core from the volume. Were the core included, there would be an additional 4\% of mass, which is less than the error quoted. This mass is similar to the average gas mass contained
within 50 kpc of early-type galaxies \citep{2013Anderson}, despite
being contained within a radius roughly half that
distance. Furthermore, K-band measurements from 2MASS give
  $L_{\rm K} = 1.6 \times 10^{10} L_{\odot}$ for 3C\,386. This is below
  the average range for early-type galaxies in Table 4 of
  \citet{2013Anderson}, and so we would expect even less gas mass in
  3C\,386, around $2 \times 10^{8} M_{\odot}$ (based on \citealt{2015Su}).  The results suggest that 3C\,386's gas
mass has been enhanced, presumably by gas originating in the group. 



In Table \ref{table:xrayprop2} we estimate some physical parameters
extrapolated from the observed group gas at roughly the distances of
the inner and outer belt. These values are derived from an
extrapolation to small radii using parameters from the radial profile
discussed in Section \ref{sec:gg}.

\begin{table*}
	\centering
	\begin{tabular}{c c c c c c}
		\hline
		$r$ (kpc) & $n_{\rm p}$ (m$^{-3}$) & $P$ ($10^{-13}$ Pa) & $M(< r)\, (M_{\odot})$ & $S$ & $S_{\rm belt}$ \\ 
		& & & & (keV cm$^{2}$) & (keV cm$^{2}$) \\
		\hline
		0 & $610^{+2000}_{-260}$ & $3.1^{+13}_{-1.3}$ & - & $190^{+600}_{-50}$ & - \\
		10 & $600^{+2000}_{-250}$ & $3.0^{+13}_{-1.3}$ & $1.5^{+2.3}_{-1.4}\times10^{8}$ & $200^{+600}_{-50}$ & $50^{+10}_{-15}$ \\
		21 & $560^{+1800}_{-240}$ & $2.8^{+12}_{-1.2}$ & $1.4^{+2.1}_{-1.3}\times10^{9}$ & $200^{+600}_{-40}$ & $130^{+40}_{-30}$ \\
		60 & $320^{+1100}_{-140}$ & $1.6^{+6.9}_{-0.7}$ & $2.3^{+3.5}_{-2.2}\times10^{10}$ & $300^{+1000}_{-70}$ & - \\
		\hline
		\end{tabular}
	  \caption{Density, pressure and pseudo-entropy (estimated using $kT/n_{\rm p}^{2/3}$) of the group gas extrapolated to radii appropriate for the inner belt (10 kpc) and outer belt (21 kpc). The inner and outer gas belt pseudo-entropy are also given.}
	\label{table:xrayprop2}
\end{table*}

Using density and temperature we can estimate cooling timescales,
$\tau_{\rm cool}$. Based on the cooling curves calculated by \citet[their Figure 5]{2006Worrall}, we
estimate $\tau_{\rm cool} \sim 5$ Gyr for the gas belt in 3C\,386.

\subsection{Physical parameters of the lobes}
The spectra of the X-ray lobes are well described by an absorbed power law (Table \ref{table:lobefit}). As described in Section \ref{sec:speclobes} the agreement between the X-ray and radio spectral indices is consistent with the interpretation of the X-ray power-law component originating from inverse-Compton (IC) emission from a population of electrons which are also producing synchrotron radio emission.

Using the volumes estimated in Section \ref{sec:speclobes} and referring to \citet{2006Worrall}, we find a minimum energy magnetic field $B_{\rm me} = 0.67$ nT for the northern lobe, and $B_{\rm me} = 0.65$ nT for the southern lobe. These were estimated assuming a filling factor of unity, no relativistic protons and no relativistic bulk motions. Minimum-energy pressure, $P_{\rm me}$, is also given in Table \ref{table:minen}.

 \begin{table}
	\centering
	\begin{tabular}{l c c}
		\hline
		Parameter & North lobe & South lobe \\
		\hline
		$B_{\rm me}$ ($10^{-10}$ T) & 6.7 & 6.5 \\
		$P_{\rm me}$ ($10^{-13}$ Pa) & 1.3 & 1.2 \\
		$B_{\rm iC}$ ($10^{-10}$ T) & 2.7 & 1.8 \\
		$P$ ($10^{-13}$ Pa) & 3.5 & 5.9 \\
		$u_{\rm B}$ ($10^{-13}$ J m$^{-3}$) & 0.3 & 0.1 \\
		$u_{\rm p}$ ($10^{-13}$ J m$^{-3}$) & 10.3 & 17.6 \\
		\hline
		\end{tabular}
	\caption{Physical parameters calculated based on the X-ray
          emission from the lobes of 3C\,386. $u_{\rm B}$ and $u_{\rm p}$ correspond to the energy density in the magnetic field and the energy density in particles respectively.}  
	\label{table:minen}
\end{table}

We also calculate the magnetic field and pressures in the lobes
assuming all X-ray emission from the region is produced by the IC
mechanism. Using this method we find lower magnetic fields in the
lobes, with $B_{\rm iC} = 0.27$ nT in the north lobe and $B_{\rm iC} = 0.18$
nT in the south lobe. The pressures, $P$, in the lobes are found to be
3 and 5 times larger than those implied by the minimum-energy
assumption. A discrepancy of a factor $\sim3$ between $B_{\rm me}$ and
$B_{\rm iC}$ for IC-emitting radio lobes is common \citep{2005Croston}.


\subsection{Model}
We have argued that there is too much gas in the belt for it to have originated entirely in the host galaxy. We suggest instead that it is being formed by a buoyancy-driven inflow from the gas atmosphere of the group on to a surviving cold core remnant of the pre-AGN atmosphere.

\begin{figure}
	\centering
	\includegraphics[width=0.99\columnwidth]{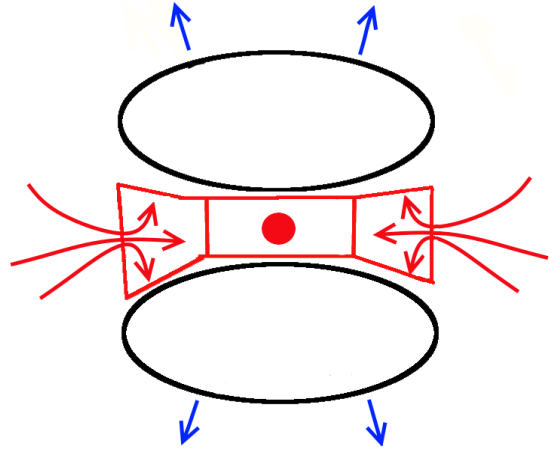}
	\caption{Once the radio source has become passive, the buoyant
          lobes rise through the ambient medium. Their motion through
          the group gas causes a `buoyancy-driven inflow' of gas,
          which forms the outer gas belt. This inflow is directed towards the relic cool core of 3C\,386.}
	\label{fig:models}
\end{figure}

When the AGN of 3C\,386 was at the height of its activity, it would
have been feeding the radio lobes with jets, causing the lobes to
expand into the intragroup gas. The AGN now appears to be driving no
significant jets into the lobes, although it continues to show a low
level of activity. The lobes are now rising buoyantly through the group medium away from the host galaxy, at a fraction of the sound speed
in the group gas ($c_{\rm s} \sim$600 km s$^{-1}$). As the lobes rise, gas
from the group atmosphere must flow back between the lobes, and we
interpret the outer gas belt as the re-accumulated group atmosphere that has
been developed by this flow. This is illustrated in Figure
\ref{fig:models}.

The temperature structure of the belt supports this interpretation.
The outer belt is hotter than the inner belt, with a temperature of
about 1.7 keV, similar to that of the group gas.  In the
inner part of the belt the gas is cooler, at about 0.6 keV. The
cooling time in the belt gas is far longer than the flow time from the
edge to the centre of the belt (5 Gyr, compared to $\sim$50 Myr), so it is not cooling that causes the central temperature to be lower. It is possible that the inner belt gas is representative of the initial state of the group gas. Its low pseudo-entropy implies it is old, unheated group gas with a small ($\leq$10\%) component of host galaxy gas. This is also supported
by the morphology of the X-ray emission in the outer belt region.  The density structure
argues against the belt having arisen from the lobes acting on a
pre-existing gas structure, as if this were the case, the inner belt
would be denser than the outer.

\begin{figure}
	\centering
	\includegraphics[width=0.99\columnwidth]{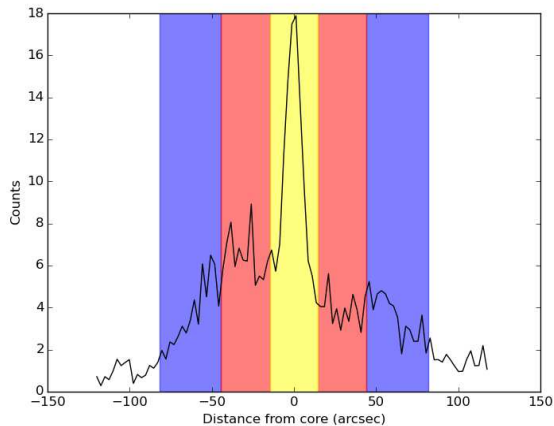}
	\caption{Background subtracted, exposure corrected counts as a
          function of distance from the core across a 20\arcsec \,
          slice of the gas belt and its surrounding group medium. Colored region indicates the extent of the gas belt region shown in Figure \ref{fig:mos2beltreg}, where blue corresponds to the outer belt, red corresponds to the inner belt and yellow corresponds to the core.}
	\label{fig:projection}
\end{figure}

Figure \ref{fig:mos2beltreg} seems to show that the gas belt has a sharp
edge, which might not be expected from a buoyancy-driven inflow. However, a plot of the counts as a function of distance across the belt (Figure \ref{fig:projection})
reveals a more gradual change in the number of counts with distance
from the core.

There are some remaining issues with our interpretation of the gas belt. 
First, the mass of gas in the belt $\sim2 \times 10^{9} M_{\odot}$, is about 
10\% of the total mass of the group gas within a sphere encompassing the radio 
lobes. 
The efficiency of collection of group gas into the belt region is unexpectedly 
high, and would correspond to a high mass flow rate, $\sim40 M_{\odot}$ yr$^{-1}$, 
into the belt. A redshift survey of 3C\,386's group would be useful to help
improve our understanding of the extent and composition of the group as currently little is known. This may help identify whether a merger in this group helped
reshape the gas distribution.

Second, the entropy of gas in the inner belt is significantly lower than that of the group gas, but the cooling time for the belt gas is too long for it to have cooled significantly over the lifetime of the radio source. The cold gas must therefore predate the radio activity, and the high mass means that it must be residual cool gas from the group. The group is poor in galaxies, consistent with the low gas temperature in the inner belt. The higher present temperature of the group gas and outer belt can then be interpreted as resulting from drastic heating driven by 3C\,386 - the temperature difference between the inner belt and group roughly matches the heating that the lobes can provide.  At the new temperature of the group, we expect much of the gas to escape. The system was previously in hydrostatic equilibrium at $\sim0.7$ keV and is now at a temperature of at least 1.4 keV which could result in the group scale atmosphere being blown away. To test this, we calculate the total mass of the group contained within a specified radius using the following equation \citep{2006Worrall}:

\begin{equation}
\label{eqn:totmasshse}
M_{\rm tot}(< r) = \frac{3\beta kT r^{3}}{G \mu m_{\rm H}(r^{2} + r_{\rm cx}^{2})}
\end{equation}

\noindent To a radius of 80 kpc, this gives a total group mass of $6.5^{+18}_{-6.4} \times 10^{12} M_{\odot}$. This is typical for galaxy groups \citep{2000Mulchaey}. We then find a gas mass fraction of $0.01^{+0.02}_{-0.005}$. This is much smaller than an expected cosmological value of about 0.17 \citep{2015PlanckCol}. Based on the size of the lobes and the sound speed in the group atmosphere, the age of the current outburst is 50 Myr. The sound speed of the gas is roughly equal to the escape velocity of the group, and it would take around 100 Myr for all the gas to flow out of the group. From these numbers, it is possible that a substantial fraction of the gas in the group has escaped due to heating by the radio source. It is also possible that multiple outbursts have been involved. Traces of a previous outburst may be sought in a low-frequency spectral index map, though we note that existing low frequency data show a relatively flat radio spectrum. Alternatively, the total mass of the 3C\,386 group could be overestimated. Equation {\ref{eqn:totmasshse}} applies only if the system is in hydrostatic equilibrium, whereas the 3C\,386 system could be far from equilibrium if an inflow is present. If the total mass of the group is 10 times lower, so that the baryonic mass fraction is more normal, the total mass becomes of order a single galaxy mass, presumably the 3C\,386 host galaxy itself. 

If we consider the work done by the lobes and calculate the bubble enthalpy ($4PV \approx 6 \times 10^{51}$ J) and compare this to the thermal energy of all the gas out to the radius of the lobes ($E = 1.5^{+4.7}_{-1.4} \times 10^{52}$ J), we see that within errors, the values are equal. This provides a lower limit to the total energy of the outburst, although an extended supersonic phase could increase this substantially.

Finally, the outer belt appears overpressured relative to the inner belt and the radio lobes, although its entropy is close to that of the group gas. A high pressure in this region is unexpected, since the buoyant motion of the lobes should generate only small pressure imbalances and slow gas flows back towards the 3C\,386 galaxy. The geometry of the belt is uncertain. For our calculations, we have assumed that the belt is a flat disk of constant height. However, looking at Figure \ref{fig:abcdsplit}, it is apparent that the outer belt splays outwards from the inner belt. This splaying could lead to the volume of the outer belt increasing by a factor of 2. The pressure in the outer belt would then fall to $5.8^{+1.5}_{-1.4} \times 10^{-13}$ Pa, bringing the outer belt closer to pressure balance with the inner belt, with any remaining imbalance attributable to the dynamics in the system.

Additionally, we are limited by our current fit of the group gas temperature,
having had to fix this parameter at the lower bound of its fitted range. Current X-ray missions are unlikely to be able to improve the observational situation.
Future missions, such as \emph{ATHENA}, with its significantly improved sensitivity, may be needed to constrain better the properties of the group
gas.

\section{Conclusion}
We have reported measurements of the extended X-ray
emission in the low-excitation radio galaxy 3C\,386. We find that
emission from the lobes is consistent with the interpretation of the
X-ray power-law component originating from IC emission. The implied
departure from equipartition of the lobes is comparable to the range
seen in other sources.

We find that the X-ray emission from the gas belt in 3C\,386 is most
likely thermal, with an average temperature of
0.94$^{+0.04}_{-0.05}$ keV across the whole of the belt. We find that
the belt displays a clear temperature structure, with the gas closer
to the core cooler than the gas closer to the surrounding group
medium. We interpret this temperature structure as indicating that the belt was likely formed by a combination of gas components. In the outer belt a `buoyancy-driven inflow' of part of the group-gas atmosphere, caused by the buoyant rising of the radio lobes through the ambient medium once the radio source became passive, is causing hot group gas to flow towards a relic cool core. This relic cool core is formed of old group gas, predating the radio activity in 3C\,386.

It is possible that this inflow could lead to a resupply of fuel to
the AGN, caused by gas reaching the parsec-scale regions surrounding
the central black hole, which would lead to 3C\,386 resuming a high
level of radio activity in the future.

\section{Acknowledgements}
The authors thank the anonymous referee for their thorough report with many useful suggestions. This research has made use of the NASA/IPAC Extragalactic Database
(NED) which is operated by the Jet Propulsion Laboratory, California
Institute of Technology, under contract with the National Aeronautics
and Space Administration. RTD thanks the STFC for their support.

\bibliography{bib}

\label{lastpage}

\end{document}